\documentclass[12pt]{article}%
\usepackage{cite}
\usepackage{graphicx}
\usepackage{multicol}
\usepackage{amsfonts}
\usepackage{amssymb}
\usepackage{amsmath}
\usepackage{heck}
\usepackage{setspace}
\usepackage{verbatim}
\usepackage{color}
\usepackage{epsfig}
\usepackage[margin=1in]{geometry}%
\providecommand{\U}[1]{\protect\rule{.1in}{.1in}}
\numberwithin{equation}{section}

\newcommand{\ba}{\begin{eqnarray}}
\newcommand{\ea}{\end{eqnarray}}

\begin{document}

\date{November 2013}
\title{Building Blocks for Generalized \\[4mm] Heterotic/F-theory Duality}

\institution{HARVARD}{\centerline{${}^{1}$Jefferson Physical
Laboratory, Harvard University, Cambridge, MA 02138, USA}}

\institution{HARVARDmath}{\centerline{${}^{2}$Department of
Mathematics, Harvard University, Cambridge, MA 02138, USA}}

\authors{Jonathan J. Heckman\worksat{\HARVARD}\footnote{e-mail: {\tt jheckman@physics.harvard.edu}},
Hai Lin\worksat{\HARVARD, \HARVARDmath}\footnote{e-mail: {\tt
hailin@fas.harvard.edu}}, and Shing-Tung Yau\worksat{\HARVARD,
\HARVARDmath}\footnote{e-mail: {\tt yau@math.harvard.edu}}}

\abstract{In this note we propose a generalization of
heterotic/F-theory duality. We introduce a set of non-compact
building blocks which we glue together to
reach compact examples of generalized duality pairs. The F-theory building blocks
consist of non-compact elliptically fibered Calabi-Yau fourfolds
which also admit a $K3$ fibration. The compact elliptic model obtained by gluing
need not have a globally defined $K3$ fibration. By replacing the
$K3$ fiber of each F-theory building block with a $T^2$, we reach building blocks in a heterotic
dual vacuum which includes a position dependent dilaton and three-form flux. These
building blocks are glued together to reach a heterotic flux background.
We argue that in these vacua, the gauge fields of the
heterotic string become localized, and remain dynamical even when
gravity decouples. This enables a heterotic dual for the hyperflux
GUT breaking mechanism which has recently figured prominently in
F-theory GUT models. We illustrate our general proposal with
some explicit examples.}

\maketitle

\tableofcontents

\enlargethispage{\baselineskip}

\setcounter{tocdepth}{2}

\newpage

\section{Introduction \label{sec:INTRO}}

One of the remarkable insights from the discovery of string dualities is that
non-perturbative physics in one duality frame can sometimes have a very simple
and exact geometric description in another duality frame. In the context of
string compactification, this is the statement that two seemingly very
different compactifications may nevertheless specify identical low energy
effective field theories.

A notable example of this type is the six-dimensional duality between
heterotic strings on $T^{4}$ and type II\ strings on a $K3$ surface
\cite{Duff:1994zt, Witten:1995ex, Sen:1995cj, Harvey:1995rn, Kachru:1995wm},
and its eight-dimensional lift to heterotic strings on $T^{2}$ and F-theory on
an elliptic $K3$ surface \cite{VafaFTHEORY, MorrisonVafaI, MorrisonVafaII,
BershadskyPLUS}. This duality can also be extended to lower dimensional
theories by fibering each side over a common base manifold.

Of course, there are a broad class of F-theory and heterotic vacua which do
not have such a dual. Indeed, recently there has been renewed interest in
F-theory as a starting point for building Grand Unified Theories (GUTs)
\cite{BHVI, BHVII, DWI, DWII} (for recent reviews see for example \cite{HVLHC,
Heckman:2010bq, Weigand:2010wm, Maharana:2012tu, Wijnholt:2012fx}). An
important aspect of these F-theory GUT models is that there is no dual
heterotic Calabi-Yau compactification \cite{BHVII, DWII}.

The reason for the absence of such a dual can be traced to the mechanism of
breaking the higher dimensional GUT group $SU(5)_{GUT}$ down to the Standard
Model gauge group. In F-theory, GUT breaking can be realized through a local to
global topological condition because gauge theory degrees of freedom are
trapped on a seven-brane. This makes it possible for an abelian flux valued in
$U(1)_{Y}$ to topologically decouple from all bulk axions \cite{BHVII, DWII}.

By contrast, in heterotic Calabi-Yau compactification, bulk axions and 10D
gauge fields propagate over the same geometry, so no topological decoupling
is available. Indeed, any such flux breaking generates a string scale mass for
the gauge boson \cite{WittenU(1)} (see also \cite{BlumenhagenMosterWeigandI,
BlumenhagenMosterWeigandII, TatarWatariU1}). Rather, in heterotic compactification on a Calabi-Yau, GUT\ breaking is
accomplished by a choice of discrete Wilson line valued in the $U(1)_{Y}$
hypercharge subgroup of $SU(5)_{GUT}$. This puts specific restrictions on the
choice of the Calabi-Yau via its fundamental group. Recent model
building efforts in heterotic theory with abelian fluxes have been considered for example
in \cite{Anderson:2012yf}.

In spite of these distinctions, there is a striking formal similarity between
the gauge theory sectors of heterotic strings and local F-theory models
defined by gauge theory on a seven-brane. So, while there are potential
discrepancies at the level of the closed string/gravitational sector, there is
a close correspondence at the level of individual gauge group factors.

Motivated by these considerations, in this note we propose a generalized
version of heterotic/F-theory duality which covers a broader
class of vacua. The basic idea will be to consider F-theory on a collection
of \textit{non-compact} elliptic Calabi-Yau fourfolds $X_{L}$, $X_{\text{mid}%
}$, and $X_{R}$, and to glue these back together along Calabi-Yau divisors to
produce a \textit{compact} elliptic model,%
\begin{equation}
X_{\text{F-th}}=X_{L}\cup_{Y_{L}}X_{\text{mid}}\cup_{Y_{R}}X_{R}.
\end{equation}
Here, $Y_{L}$ is a Calabi-Yau divisor common to $X_{L}$ and $X_{\text{mid}}$,
and $Y_{R}$ is a Calabi-Yau divisor common to $X_{R}$ and $X_{\text{mid}}$.
Each factor $X_{(i)}$ is a $K3$ fibration over a four-manifold $S_{(i)}$, so
each threefold base is a $\mathbb{P}^{1}$ fibration over $S_{(i)}$. However,
we only demand a section for the elliptic fibration, and do not require a global $K3$
fibration for $X_{\text{F-th}}$.\footnote{One might
ask why we do not consider the comparatively simpler case of six-dimensional
heterotic and F-theory vacua. The reason is that if we have an F-theory model
with a complex twofold base which also has a $\mathbb{P}^{1}$ fibration over a
real dimension two manifold, then this fibration also has a section. For example, if the
fibration is over $\mathbb{P}^{1}$ then we are dealing with F-theory with base
a Hirzebruch surface, which is a case covered by the standard duality.}

\begin{figure}[ptb]
\begin{center}
\includegraphics[
height=1.3638in,
width=4.6812in
]{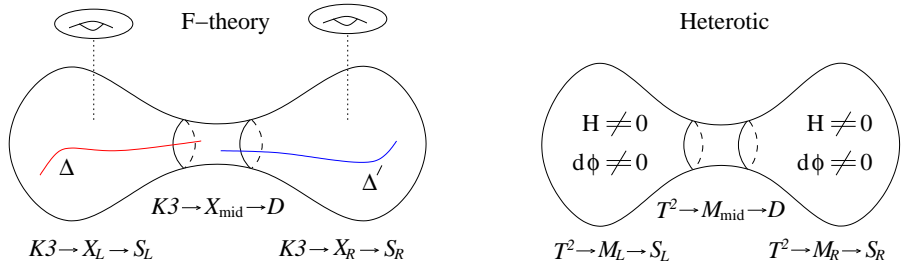}
\end{center}
\caption{Depiction of the non-compact building blocks used to generate
F-theory/heterotic pairs. On the F-theory side, we have non-compact building
blocks each given by a $K3$ fibration over a four-manifold. On the heterotic
side, these $K3$ fibers are replaced by $T^{2}$ fibers. In the middle region
$M_{\text{mid}}$ of the heterotic geometry, the string coupling becomes big,
localizing the gauge field degrees of freedom to the gluing regions between
$M_{L}$ and $M_{\text{mid}}$, and between $M_{R}$ and $M_{\text{mid}}$.}%
\label{configurationnew}%
\end{figure}

In the heterotic dual, we have building blocks $M_{L}$,
$M_{\text{mid}}$, and $M_{R}$, which are to be glued together both at the
level of the geometry,%
\begin{equation}
M_{\text{het}}=M_{L}\cup_{D_{L}}M_{\text{mid}}\cup_{D_{R}}M_{R},
\end{equation}
as well as through the profile of the dilaton and three-form flux. The divisor
$D_{L}$ is common to $M_{L}$ and $M_{\text{mid}}$, and the divisor $D_{R}$ is
common to $M_{R}$ and $M_{\text{mid}}$. The non-compact
factors $M_{(i)}$ are obtained by starting with the base of an F-theory building block,%
\begin{equation}
\text{F-th}\text{: \ \ }\mathbb{P}^{1}\rightarrow B_{(i)}\rightarrow
S_{(i)},
\end{equation}
and replacing the $\mathbb{P}^{1}$ fiber with a $T^2$,
\begin{equation}
\text{Het}\text{: \ \ }T^{2}\rightarrow M_{(i)}\rightarrow
S_{(i)}.
\end{equation}
See figure \ref{configurationnew} for a depiction of the F-theory and heterotic sides of the duality.

It is helpful to view these non-compact building blocks as coming from compact
geometries which by themselves would define inconsistent Minkowski vacua. For
example, on the F-theory side of the correspondence, we work with a $K3$
fibration over $S$, where the total space is not Calabi-Yau. We reach a
non-compact Calabi-Yau by deleting a divisor from this fourfold. On the
heterotic side, we start with a compact threefold of positive curvature with a
position dependent dilaton and three-form flux switched on. In both cases, we
arrive at a ten-dimensional spacetime consistent with the supergravity
equations of motion by deleting an appropriate subspace.

An important feature of our proposal is that the F-theory side remains purely
geometric. On the heterotic side, we instead have a non-trivial profile for
the background fields, including the dilaton and three-form flux. Indeed, in
the absence of the analogue of Yau's theorem for Calabi-Yau compactification,
it has proven necessary to construct on a case by case basis heterotic flux
backgrounds of the type proposed in \cite{Strominger:1986uh}. For recent work
on heterotic flux compactifications, see for example \cite{Fu:2006vj,
Becker:2006et, Fu:2008ga, Becker:2009df, AndreasHflux}.

Our proposal uncovers a number of novel physical mechanisms in heterotic
theory. First of all, because the volume of the $\mathbb{P}^{1}$ fiber on the
F-theory side is not constant, the profile of the heterotic dilaton will be
position dependent. From this we see that there will be geometrically
separated regions of weak coupling and strong coupling. This leads to pockets
where the ten-dimensional gauge fields effective localize, which in turn points the way to a
hyperflux GUT group breaking mechanism for heterotic strings. The key
difference from standard Calabi-Yau compactification of the heterotic string
is the presence of a position dependent dilaton and three-form flux.

To illustrate the general contours of our proposal, we also
present some examples, focussing mainly on the case of F-theory with a
threefold base $\mathbb{P}^{3}$. The key feature of this example is that
$\mathbb{P}^{3}$ is the twistor space for $S^{4}$, that is, we have a
$\mathbb{P}^{1}$ fibration over $S^{4}$. In spite of this, there is no
standard heterotic dual because the elliptic fibration does not extend to a
global $K3$ fibration. By taking a stable degeneration limit, however, we will
arrive at two eight-manifolds, each of which is given by a $K3$ fibration over
$S^{4}$. For each of these building blocks, we get a heterotic dual, which we
piece together to form a new heterotic/F-theory pair. This F-theory model is
consistent with the hyperflux mechanism \cite{BHVII}. Applying our duality, we
translate this mechanism over to the heterotic side of the correspondence.

We believe that the arguments presented here provide strong evidence for the existence of
a new class of heterotic/F-theory dualities. However, we leave more detailed checks
of our proposal for future work.

The organization of the rest of this paper is as follows. First, in section
\ref{sec:REVIEW} we briefly review some aspects of the standard duality. In
section \ref{sec:FBLOCKS} we turn to the F-theory building blocks, and in
section \ref{sec:HBLOCKS} we determine their heterotic duals. In
section \ref{sec:GRAVITY}, we show how to glue together these non-compact
building blocks to reach compact models with dynamical
gravity. This also leads us to a heterotic dual of the hyperflux mechanism of
F-theory GUTs. After this, in section \ref{sec:EXAMP} we
turn to some examples, and in section \ref{sec:CONCLUDE} we present our
conclusions and directions for future investigation. Some additional details
on elliptic fourfolds with a $\mathbb{P}^{3}$ base are collected in an Appendix.

\section{The Standard Duality \label{sec:REVIEW}}

In preparation for our later analysis, in this section we briefly review the
standard heterotic/F-theory duality (see for example \cite{VafaFTHEORY,
MorrisonVafaI, MorrisonVafaII}). In particular, we emphasize those aspects of
the duality which we will later aim to generalize. Recall that the standard
duality involves heterotic strings compactified on $T^{2}$, which is dual to
F-theory compactified on an elliptically fibered $K3$ surface. We can extend
this duality fiberwise to reach lower-dimensional dualities for the geometries,
\begin{align}
\text{F-th} &  \text{: \ \ }K3\rightarrow X\rightarrow S,\label{firstline}\\
\text{Het} &  \text{: \ \ }T^{2}\rightarrow M\rightarrow S,
\end{align}
where $S$ is taken to be a K\"{a}hler surface, and $X$ and $M$ respectively
define an elliptically fibered Calabi-Yau fourfold and threefold, each with a
section. The F-theory model is also given by an elliptic fibration,
\begin{align}
\text{F-th}  & \text{: \ \ \ }T^{2}\rightarrow X\rightarrow B,\\
\text{Base}  & \text{: \ \ \ }\mathbb{P}_{f}^{1}\rightarrow B\rightarrow S,
\end{align}
where the base $B$ is itself a $\mathbb{P}_{f}^{1}$ fibration over $S$. The
$\mathbb{P}_{f}^{1}$ fiber is also the base for the elliptic $K3$ of line
(\ref{firstline}). In the standard duality, this $\mathbb{P}_{f}^{1}$
fibration has a section.

Now, an important feature of this duality is the repackaging of the vector
bundle degrees of freedom of the heterotic theory in purely geometric terms in
F-theory. This is simplest to arrange in the stable degeneration limit, where
we take the $K3$ fiber, and let it degenerate to a pair of del Pezzo nine
($dP_{9}$) surfaces. Each one of these $dP_{9}$'s can carry an $E_{8}$ singularity.

Since we will be generalizing the stable degeneration limit, let us now
explain in more detail how this works. To begin, we consider a del Pezzo nine
surface. It is described by an elliptic fibration $T^{2}\rightarrow
dP_{9}\rightarrow\mathbb{P}^{1}$. The minimal Weierstrass model for this is
\begin{equation}
y^{2}=x^{3}+f_{4}x+g_{6},
\end{equation}
where $f_{4}$ and $g_{6}$ are degree four and six homogeneous polynomials in
the homogeneous coordinates of the base $\mathbb{P}^{1}$. This geometry should
be viewed as \textquotedblleft half of a K3\textquotedblright, since for a
$K3$ surface we would take degrees $8$ and $12$ for $f$ and $g$, respectively.

Now, although $dP_{9}$ is not Calabi-Yau, we can manufacture a non-compact
Calabi-Yau by deleting an appropriate subspace from this geometry. To see how
this works, recall that $dP_{9}$ can also be viewed as a $\mathbb{P}^{2}$
blown up at nine points. The anti-canonical class for $dP_{9}$ is,%
\begin{equation}
-K_{dP_{9}}=3H-(E_{1}+...+E_{9}),
\end{equation}
where $H$ is the hyperplane class of the $\mathbb{P}^{2}$, and $E_{i}$ are the
exceptional divisors. This divisor class also defines an elliptic curve in the
$dP_{9}$ geometry. The holomorphic two-form of $dP_{9}$ has a pole along this
elliptic curve. By deleting it, we reach a non-compact Calabi-Yau.

We can now glue this non-compact Calabi-Yau with another copy to produce a
$K3$ surface. To do this, we take a limit for the Calabi-Yau metric where the
base $\mathbb{P}^{1}$ stretches out to a long cylinder. At one end of the
cylinder, we have our original $dP_{9}$ with the elliptic curve subtracted. At
the other end, we take another copy of $dP_{9}$ with an elliptic curve
deleted. Gluing the two pieces together, we get a new geometry. This is the
stable degeneration limit for the $K3$. In the dual heterotic string picture,
this degeneration limit maps to the strongly coupled regime of the heterotic
string, that is, heterotic M-theory with two $E_{8}$ nine-branes.

This limit allows a number of precision checks of the duality. For example,
the geometric moduli of a $dP_{9}$ factor maps to the vector
bundle moduli of a given $E_{8}$ factor. Additionally, the leading order
profile for the heterotic dilaton is simply a constant, being set by the ratio
of volumes for the fiber $\mathbb{P}_{f}^{1}$ to the base $S$ in string units.

It is also interesting to consider perturbations away from this limit. For
example, a mild position dependence in the heterotic dilaton corresponds to a
position dependent profile for $\mathrm{Vol}(\mathbb{P}_{f}^{1})/\mathrm{Vol}%
(S)$, where $S$ here denotes the base in the standard duality. Additionally,
we can see that at least at a qualitative level, the three-form flux of the
heterotic theory converts, by the chain of dualities
Het$_{E_{8}\times E_{8}}\leftrightarrow$ Het$_{SO(32)}\leftrightarrow$ Type I $\leftrightarrow$ F-th
to a non-trivial profile in F-theory for the RR one-, three- and five-form
fluxes, depending on how many legs of the three-form flux are on the $T^{2}$ fiber.

In the cases of interest to us in this paper, we focus on Calabi-Yau
compactification of F-theory. This means we shall exclude the possibility of a
five-form flux on the F-theory side, and so on the heterotic side, the B-field
must have at least one leg along a $T^{2}$ fiber. Though we will give a more
concrete proposal for the presence of such fluxes later, it is instructive to
study how activating such modes shows up in the standard duality.

In eight dimensions, a B-field with two legs on the $T^{2}$ of the heterotic
theory corresponds to activating a specific complex modulus in the $T^{2}$
used to glue the two $dP_{9}$ factors together on the heterotic side,
\begin{equation}
y^{2}=x^{3}+\alpha x+\beta,
\end{equation}
for $\alpha$ and $\beta$ constant. One combination corresponds to the complex
structure of the $T^{2}$, and since we have two marked points on the base
$\mathbb{P}^{1}$, the other combination fixes the complexified K\"{a}hler
modulus. Fibering this over another geometry, we see that in general, $\alpha$
and $\beta$ will pick up a position dependent profile. On the heterotic side,
we get a position dependent B-field, which we interpret as a possibly
non-trivial three-form flux.

\section{F-theory Building Blocks \label{sec:FBLOCKS}}

In this section we describe our procedure for building up more general
heterotic/F-theory pairs. To this end, here we show how to manufacture
non-compact elliptic Calabi-Yau fourfolds which also admit a $K3$ fibration.
Our plan will be to glue these non-compact building blocks together to get
compact elliptic Calabi-Yau fourfolds. Tracking each component through to a
heterotic dual, we shall then get a generalization of the standard duality.

\subsection{Geometric Gluing \label{ssec:GLUE}}

In preparation for our later discussion, let us now review some
aspects of how to glue together manifolds along a subspace. To frame our
discussion, suppose we are given two complex varieties $X_{(1)}$ and $X_{(2)}
$ of dimension $n$ which both contain some divisor $Y$. Then, we can form a
new topological space,%
\begin{equation}
X_{(1)}\cup_{Y}X_{(2)},
\end{equation}
by first deleting $Y$ from each region, and then identifying points in the two
deleted regions. Deleting such a divisor will also alter the metric near the
deleted locus. To illustrate, suppose $X_{(1)}$ is not Calabi-Yau, i.e. the
holomorphic $n$-form has some pole along a divisor $Y$. Then, we reach a
non-compact Calabi-Yau by deleting this region from $X_{(1)}$, that is, we get
a non-compact Calabi-Yau $X_{(1)}\backslash Y$. Moreover, the divisor $Y$ is
itself Calabi-Yau, and the divisor class of $Y$ is the anti-canonical class of
$X$. By the adjunction formula, the canonical class of $Y$ is,
\begin{equation}
K_{Y}=(K_{X}+[Y])|_{Y}, \label{KDadjunct}%
\end{equation}
so since $K_{X}=-[Y]$, we see that $Y$ is in fact Calabi-Yau.

More generally, deleting a divisor $Y$ from a variety $X$ alters the profile
of the metric on the non-compact space. This makes it possible to generate
non-compact Calabi-Yau geometries from compact positive curvature geometries
\cite{TY1, TY2, Bando_Kobayashi}. Consider a compact manifold $X$ with
positive first Chern class $c_{1}(X)>0$, and a divisor $Y$ of the manifold $X
$ with its divisor class $[Y]$. There is an exact sequence,
\begin{equation}
\mathbb{Z\cdot\lbrack}Y]\rightarrow\mathrm{Pic}(X)\rightarrow\mathrm{Pic}%
(X\backslash Y)\rightarrow0,
\end{equation}
which means that for the inclusion map $i:X\backslash Y\rightarrow X,~$the
kernel of the Picard group of $X$ under the pull-back $i^{\ast}$, is the
integer multiple of the divisor class of $Y$.

In the cases of interest to us in this paper, the anti-canonical bundle will
be a multiple of some divisor $Y$, which is $-K_{X}=r[Y]$ for $r\geq1$. In the
case $r=1$, we have already argued that the non-compact space $X\backslash Y$
is Calabi-Yau, and the divisor $Y$ is also Ricci flat.

For $r>1$, the adjunction formula implies$~c_{1}(Y)=(-K_{X}-[Y])|_{Y}%
=(r-1)[Y]|_{Y}>0$, so the divisor $Y$ has positive curvature for $r>1$. Let
$\xi=-K_{X}-r[Y]$ represent the first Chern class. If $Y$ has a K\"{a}hler
metric with K\"{a}hler-form $\omega_{Y}$, and Ricci-form $\mathrm{Ric}%
(\omega_{Y}),$ and
\begin{equation}
\mathrm{Ric}(\omega_{Y})=(r-1)\omega_{Y}+\xi,
\end{equation}
then there is a complete K\"{a}hler metric $g_{\xi}$ on $X\backslash Y$ with
Ricci curvature form $\xi$. If, in addition, $Y$ is K\"{a}hler-Einstein, then
$\mathrm{Ric}(\omega_{Y})=(r-1)\omega_{Y}>0$, so it implies that $\xi=0$, and
the complement $X\backslash Y$ has a complete Ricci-flat K\"{a}hler metric.

\subsection{Building Blocks with $K3$ Fibers}

In this section we introduce the F-theory building blocks defined by a
non-compact elliptic Calabi-Yau which also admits a $K3$ fibration. Our plan
will be to glue these pieces together to reach a compact model.

Our starting point is $X$, a positive curvature K\"{a}hler fourfold with an
elliptic fibration with section which also admits a $K3$ fibration,
\begin{align}
T^{2}  &  \rightarrow X\rightarrow B,\\
K3  &  \rightarrow X\rightarrow S,
\end{align}
where $B$ is a complex threefold and $S$ is a four-manifold. We assume that
the elliptic fibration has a section. However, we do not assume a section for
the $K3$ fibration. By construction, the base of the elliptic fibration is a
complex threefold with a $\mathbb{P}^{1}$ fibration,%
\begin{equation}
\mathbb{P}_{\text{fiber}}^{1}\rightarrow B\rightarrow S.
\end{equation}
An example of this type we return to later will be $B=\mathbb{P}^{3}$ and
$S=S^{4}$.

Since we have assumed $X$ has positive curvature, the holomorphic four-form
will have a pole along some divisor $Y\subset X$. We reach a non-compact
Calabi-Yau geometry by deleting $Y$ from the geometry. Following up on our
general discussion in subsection \ref{ssec:GLUE}, $Y$ is a Calabi-Yau threefold.

Because we have assumed the existence of an elliptic fibration with section,
the divisor $Y$ is an elliptically fibered Calabi-Yau threefold with section.
The base of the fibration is a divisor $D$ contained in $B$,
\begin{equation}
T^{2} \rightarrow Y\rightarrow D.
\end{equation}
We delete $Y$ from $X$, and $D$ from $B$ to get non-compact geometries,
\begin{equation}
X_{L}=X\backslash Y\text{ \ \ and \ \ }B_{L}=B\backslash D.
\end{equation}
Upon performing this excision, we reach a non-compact elliptically fibered Calabi-Yau fourfold
$X_{L}$ with base $B_{L}$. So in other words, we get an F-theory model on
the non-compact background $\mathbb{R}^{3,1} \times X_{L}$.

Let us now study the low energy effective field theory defined by this
non-compact F-theory geometry. We have an elliptic fibration over a
non-compact base, with minimal Weierstrass model,%
\begin{equation}
y^{2}=x^{3}+fx+g,
\end{equation}
where $f$ and $g$ are sections of $K_{B_L}^{-4}$ and $K_{B_L}^{-6}$,
respectively. The discriminant locus is the zero set of:
\begin{equation}
\Delta=4f^{3}+27g^{2},
\end{equation}
which is a section of $K_{B_L}^{-12}$. We shall also refer to the
discriminant locus as $\Delta$, and specific components of it by $\Delta_{i}$.

Now in F-theory, we associate the components of the discriminant locus with
subspaces of $B$ wrapped by seven-branes. For appropriate singularity types on
a component $\Delta_{i}$, we get a seven-brane with a gauge group $G_{i}$. In
the low energy effective field theory in the uncompactified directions, the
value of the gauge coupling is (in Einstein frame) proportional to the volume
of $\Delta_{i}$, that is,
\begin{equation}
\frac{1}{g_{(i)}^{2}}\propto\text{Vol}(\Delta_{i}).
\end{equation}
These seven-branes are non-dynamical because in the non-compact geometry the divisors $\Delta_i$
have infinite volume. Indeed, inside of the base $B$, the divisor $D$ and
$\Delta_{i}$ intersect along a curve, and so upon deleting
$D$, $\Delta_{i}$ becomes non-compact.

To make the seven-branes dynamical, but remain in a non-compact geometry, we
can glue back in the deleted components of the discriminant locus. In more
detail, we now construct an asymptotic Calabi-Yau geometry which glues into
$Y$. In the vicinity of the deleting locus, it is given by the product
$Y\times\mathbb{C}^{\ast}$. However, since we will need to glue this geometry
to another compact component, we shall allow a fibration of $Y$ over
$\mathbb{C}^{\ast}$ away from the gluing region. We view the cylinder
$\mathbb{C}^{\ast}$ as a $\mathbb{P}_{\text{cyl}}^{1}$ with two marked points
deleted, so we can introduce the \textquotedblleft middle
geometry\textquotedblright,\
\begin{equation}
Y\rightarrow X_{\text{mid}}\rightarrow\mathbb{P}_{\text{cyl}}^{1}.
\end{equation}
At the south pole of the $\mathbb{P}_{\text{cyl}}^{1}$ we will be gluing into
the geometry $X_{L}$. At the north pole of the $\mathbb{P}_{\text{cyl}}^{1}$
we will instead glue into a new geometry $X_{R}$. The $X_{R}$ is a manifold
similar to $X_{L}$, and can also be constructed by $X_{R}=X\backslash Y$%
.$~$Since $Y$ is itself an elliptic fibration over $D$, we see that
$X_{\text{mid}}$ also defines a consistent F-theory model with threefold base
$B_{\text{mid}}$ given by fibering the divisor $D$ over this $\mathbb{P}%
_{\text{cyl}}^{1}$,%
\begin{equation}
D\rightarrow B_{\text{mid}}\rightarrow\mathbb{P}_{\text{cyl}}^{1}.
\end{equation}

Up to this point, we have kept the choice of the above fibration arbitrary.
However, to set up a building block with a heterotic dual, we need to also
have a $K3$ fibration over a complex twofold base. One simple way to arrange
this is to further restrict $B_{\text{mid}}$ to be a product manifold
$D\times\mathbb{P}_{\text{cyl}}^{1}$, in which case $X_{\text{mid}}~$is%
\begin{equation}
T^{2}\rightarrow X_{\text{mid}}\rightarrow D\times\mathbb{P}_{\text{cyl}}^{1}.
\end{equation}
We shall mainly focus on this case, since it holds more generally. Another
possibility is to assume that $D$ is a Hirzebruch surface. In section
\ref{sec:EXAMP} we shall also consider the case of the non-trivial fibration
$D \rightarrow B_{\text{mid}} \rightarrow \mathbb{P}^1$, where we relate the degree(s) of the fibration
to background instanton numbers for the $E_{8}$ factors of the heterotic dual.

Having introduced two elliptic Calabi-Yau fourfolds which share a common
region $Y$, we can glue these back together along $Y$ to produce another
non-compact Calabi-Yau
\begin{equation}
X_{L,\text{mid}}\equiv X_{L}\cup_{Y}X_{\text{mid}}.
\end{equation}
In the base of each elliptic fibration, we are gluing along a common divisor
$D$,%
\begin{equation}
B_{L,\text{mid}}\equiv B_{L}\cup_{D}B_{\text{mid}}.
\end{equation}
Introducing the dualizing sheaf $K_{L,\text{mid}}$ for $B_{L,\text{mid}}$, we
see that the minimal Weierstrass model extends as well, with $f$, $g$ and the
discriminant, sections of $K_{L,\text{mid}}^{-4}$ and $K_{L,\text{mid}}^{-6}$,
and $K_{L,\text{mid}}^{-12}$, respectively.

Now, the whole point of introducing the extra gluing by $X_{\text{mid}}$ was
to ensure that our seven-branes from the $X_{L}$ model would now be compact.
To see that this has happened, observe that each component $\Delta_{i}$ of the
discriminant locus for $X$ intersects $D$ along a curve $\Sigma_{i}\subset D$.
So, we see that in $B_{\text{mid}}$, these pieces have been added back in:
Inside of $B_{\text{mid}}$, these seven-branes sit at a point of
$\mathbb{P}_{\text{cyl}}^{1}$, and wrap the curve $\Sigma_{i}\subset D$. Note
that gravity is still decoupled because $B_{L,\text{mid}}$ is non-compact and
has infinite volume.

So far, we have focussed on the geometry of the elliptic fibration. We can
also see how this gluing works when we view $X$ as a $K3$ fibration over $S$.
There is a subtlety here, because we are not assuming the existence of a
section for the $K3$ fibration. This means $S$ need not exist as a
submanifold in either $B$ or $X$. However, we can still consider the image of $D$ under the
pushforward $\pi: B \rightarrow S$. This image defines a real two-dimensional subspace $P \subset S$. Deleting
$D$ from $B$ then means we must delete $P$ from $S$, so the geometry $X_L = X \backslash Y$ is also
a $K3$ fibration over the four-manifold,
\begin{equation}
S_L = S \backslash P. \label{Sleft}
\end{equation}
After deleting $P$ from $S$, the $\mathbb{P}^1_{\text{fiber}}$ fibration over $S_L$ has a section.
With respect to the Calabi-Yau metric on $X_{L,\text{mid}}$ we also
see that the volume of $S_L$ is finite.

Finally, although the elliptic fibration naturally extends out to
$X_{\text{mid}}$, there is no extension as a $K3$ fibration. Indeed,
inside the left region, the curve $\mathbb{P}^{1}_{\text{fiber}}$ only intersects $D$
at a finite number of points. This means that in the glued together geometry, the base $\mathbb{P}%
_{\text{fiber}}^{1}$ of the $K3$ fiber in the left region has collapsed to
zero size precisely along $P$,
where we instead glue into a new geometry. Introducing the
smoothed out Calabi-Yau fourfold after gluing, the volume of the fiber
$\mathbb{P}_{\text{fiber}}^{1}$ of the left region degenerates exponentially
as we move into the interior of the middle region. Indeed, the whole point of
our construction is that the seven-branes on $X_{L}$ remain localized, even
after gluing in $X_{\text{mid}}$.

\section{Heterotic Building Blocks \label{sec:HBLOCKS}}

In this section we convert our F-theory building blocks to heterotic duals. In
the heterotic description, the gluing will be both geometric, and will also
involve a position dependent profile for the dilaton and three-form flux.

\subsection{Geometric Components}

First, we determine the geometries for each of the heterotic regions dual to
$X_{\text{mid}}$ and $X_{L}$, which we refer to as $M_{\text{mid}}$ and
$M_{L}$. Deep in the middle region, we have a standard Calabi-Yau
compactification of the heterotic string. However, in the region $M_{L}$, we
find that the heterotic string is defined over a torsional flux background.

Consider first the heterotic dual to the F-theory geometry $X_{\text{mid}}$.
Recall that a simple way to arrange a $K3$ fibration in the middle geometry is
by restricting $B_{\text{mid}}$ to be a product manifold $D\times
\mathbb{P}_{\text{cyl}}^{1}$. In this case the middle geometry is given by a
holomorphic $K3$ fibration over a base $D$
\begin{equation}
K3\rightarrow X_{\text{mid}}\rightarrow D.
\end{equation}
So, applying the standard rules of heterotic/F-theory duality, we conclude
that there is a dual heterotic compactification on a Calabi-Yau threefold%
\begin{equation}
T^{2}\rightarrow M_{\text{mid}}\rightarrow D,
\end{equation}
where the moduli of the F-theory $K3$ fiber translate to moduli of an
$E_{8}\times E_{8}$ vector bundle. In fact, we know that this geometry
$M_{\text{mid}}$ is nothing other than the Calabi-Yau divisor $Y$, which is
also an elliptic fibration over $D$,%
\begin{equation}
M_{\text{mid}}\simeq Y\text{.}%
\end{equation}
We can also see that in the simplest case where $B_{\text{mid}}=D\times
\mathbb{P}_{\text{cyl}}^{1}$, we can, much as in \cite{MorrisonVafaI,
MorrisonVafaII}, extract the value of the heterotic dilaton in the middle
region,
\begin{equation}
\exp(-2\phi_{\text{mid}})=\frac{\text{Vol}(D)}{\text{Vol}(\mathbb{P}%
_{\text{cyl}}^{1})}.
\end{equation}

Consider next the region $M_{L}$ dual to F-theory on $X_{L}$. Again, since we
have a K3-fibration, we can replace the $K3$ by a $T^{2}$ fiber. In this case,
however, there is no guarantee that the fibration admits a section. The
appropriate heterotic dual geometry is a $T^{2}$ fibration over the base
$S_{L}$,%
\begin{equation}
T^{2}\rightarrow M_{L}\rightarrow S_{L},
\end{equation}
that is, we exchange the $K3$ fibration over $S_{L}$ for a $T^{2}$ fibration.

In fact, there is a canonical way to define this fibration, starting from the
characterization of $B$ as a $\mathbb{P}_{\text{fiber}}^{1}$ fibration over
$S$. We seek a divisor $\Gamma\subset B$ which intersects this $\mathbb{P}%
_{\text{fiber}}^{1}$ precisely four times. Assuming this has been arranged, we
can consider the branched cover over these four points, producing the
corresponding $T^{2}$. This defines a double cover $\widehat{B}\rightarrow B$.
Since $M_{\text{mid}}$ and $M_{L}$ are glued along a common divisor $D$, we
see that the manifold $M_{L}$ is also obtained by deleting $D $ from
$\widehat{B}$.

Having given a characterization of the geometry $M_{L}$, we can now see that a
number of modes in the heterotic dual description are automatically switched
on. To begin, consider the profile of the dilaton. As we remarked near
equation (\ref{Sleft}), the volume of $\mathbb{P}_{\text{fiber}}^{1}$
degenerates along the subspace $P$ of $S$. We also
know that in the standard duality, the ratio of the
fiber to base volumes controls the value of the
heterotic dilaton,
\begin{equation}
\exp(-2\phi_{L})=\frac{\text{Vol}(S_{L})}{\text{Vol}(\mathbb{P}_{\text{fiber}%
}^{1})},
\end{equation}
where here, Vol$(S_{L})$ refers to the volume of $S_{L}$, viewed as a
submanifold of $B_{L}$, where $\mathbb{P}_{\text{fiber}}^{1}$ appears as the
base in the $K3$. Since we have already argued that Vol$(S_{L})$ remains
finite, while Vol$(\mathbb{P}_{\text{fiber}}^{1})$ collapses to zero at the
gluing along $P$, we see that the dilaton $\exp(2\phi_{\text{het}})$ approaches zero
near the gluing regions. As we move away from the locus $P$, the volume of
Vol$(\mathbb{P}_{\text{fiber}}^{1})$ will also change. This means that the
heterotic dual has a position dependent dilaton.

Since the dilaton is not constant, we are not dealing with a standard
Calabi-Yau compactification of the heterotic string. Rather, we have a more
general solution to the heterotic equations of motion where backgrounds fluxes
are switched on. For example, variation of the dilatino shows that a gradient
in the dilaton correlates with the presence of a non-zero three-form flux.

Let us now discuss such heterotic flux vacua. For now, we work to leading
order in $\alpha^{\prime}$ and neglect non-perturbative corrections. Since we
still have four-dimensional $\mathcal{N}=1$ supersymmetry in flat space, we
can already assert that $M_{L}$ must be a six-dimensional complex manifold
with $SU(3)$ holonomy with respect to some torsional connection. This means we
can introduce a hermitian $(1,1)$ form $J$, and a holomorphic $(3,0)$ form
$\Omega$. Solutions to the ten-dimensional supergravity equations of motion
satisfy (see for example \cite{Strominger:1986uh, Fu:2006vj, Becker:2006et}):
\begin{align}
d(\left\Vert \Omega\right\Vert _{J}\text{ }J\wedge J)  &  =0\text{ \ \ where
\ \ }\Omega\wedge\overline{\Omega}=-i\frac{4}{3}\left\Vert \Omega\right\Vert
_{J}^{2}\text{ }J\wedge J\wedge J.\\
F^{(2,0)}  &  =F^{(0,2)}=0\text{ \ \ and \ \ }F_{mn}J^{mn}=0.\\
2i\partial\overline{\partial}J  &  =\frac{\alpha^{\prime}}{4}\left[
\text{tr}(R\wedge R)-\text{tr}\left(  F\wedge F\right)  \right]  .
\end{align}
The last equation is the heterotic anomaly cancelation condition, which can
be deferred for a non-compact model by introducing a background source
\textquotedblleft at infinity\textquotedblright. Here, the curvature $R$ is
defined with respect to the hermitian form $J$, so that tr$(R\wedge R)$ is a
$(2,2)$ form. In terms of the physical fields, we have the relations%
\begin{equation}
g_{mn}=J_{mr}I^{r}{}_{n}\text{, \ \ }H=i(\overline{\partial}-\partial)J\text{,
\ \ }e^{-2\phi_{\text{het}}}=\left\Vert \Omega\right\Vert _{J},
\end{equation}
where $I^{r}{}_{n}$ is the complex structure specified by $\Omega$, and $g$
and $H$ are respectively the metric and three-form flux.

One interesting feature of such flux vacua is that in compact models, some
moduli are automatically frozen out. For example, if we find a solution for
some choice of heterotic dilaton $\phi_{\ast}$, in general we cannot simply
shift the value of the dilaton by a constant. The reason is that the dilaton
is fixed by the value of $\left\Vert \Omega\right\Vert _{J}$, and this is in
turn fixed by the constraint $H=i(\overline{\partial}-\partial)J$.

This is all to the good because on the F-theory side, this modulus is also
frozen out: It is given by the ratio of the volume for the $\mathbb{P}%
_{\text{fiber}}^{1}$ to the volume of the four-dimensional base $S_{L}$. In
the non-compact setting, this ratio is tunable, but once we glue into the full
geometry, the absence of a section for the $\mathbb{P}_{\text{fiber}}^{1}$
fibration in the full geometry means this ratio is no longer tunable.

\subsection{Heterotic Gluing}

We now explain how to glue our heterotic building blocks together to produce a
new heterotic dual. In a certain sense this must be possible because we have
already identified a geometric prescription in F-theory.

At the level of the heterotic geometry, we are gluing along the divisor $D$
deleted from $M_{L}$ and $M_{\text{mid}}$. So, we form a new geometry via,%
\begin{equation}
M_{L,\text{mid}}\equiv M_{L}\cup_{D}M_{\text{mid}}.
\end{equation}
The central point is that we can again smooth out the metric over the gluing
locus. In addition, we also need to match the profile of the heterotic fields
across the two regions.

Let us now turn to the profile of the supergravity fields. We work in the step
function approximation, i.e. prior to smoothing out the profile of the fields
across the two regions. The main observation is that because the dual F-theory
geometry can be consistently smoothed out, similar considerations apply on the
heterotic side. Though a full analysis is beyond the scope of the present
work, it is useful to identify some qualitative aspects of how this match must work.

First of all, in the vicinity of the divisor $D$, we expect the leading order
description of the heterotic flux vacuum to be captured by the Strominger
system. The reason is that in the non-compact geometry, we can indeed tune
$\exp(2\phi_{\text{het}})$ to be arbitrarily small. However, this
approximation may in principle receive higher order $\alpha^{\prime}$ and
non-perturbative corrections as we move deep into the interior of $M_{L}$.

To illustrate how the fields look near the vicinity of $D$, consider the local
geometries for $D$ inside of $M_{\text{mid}}$, as well as inside of $M_{L}$.
This is given by the normal bundles,%
\begin{align}
N_{\text{mid}}  &  \rightarrow M_{\text{mid}}^{\text{loc}}\rightarrow D,\\
N_{L}  &  \rightarrow M_{L}^{\text{loc}}\rightarrow D.
\end{align}
We denote by $z_{\text{mid}}$ the normal coordinate for $N_{\text{mid}}$ and
$z_{L}$ the normal coordinate for $N_{L}$ so that $D$ is located at
$z_{L}=z_{\text{mid}}=0$. A solution to the heterotic equations of motion
requires matching the profiles of the fields across the two regions. For
example, in the middle region $M_{\text{mid}}$, there is now a localized
source at $z_{\text{mid}}=0$, so that the profile of the string coupling and
three-form flux is,%
\begin{align}
\frac{1}{g_{\text{het}}^{2}}  &  =\frac{1}{g_{\text{mid}}^{2}}+f_{L},\\
H_{\text{het}}  &  =H_{\text{mid}}+h_{L},
\end{align}
where $g_{\text{het}}^{2}=\exp(2\phi_{\text{het}})$ sets the strength of the
string coupling, and $f_{L}$ and $h_{L}$ vanish as $\left\vert z_{\text{mid}%
}\right\vert \ $moves away from the origin. These contributions are associated
with the correction terms from the $M_{L}$ region. Here, the entries
$\phi_{\text{mid}}$ and $H_{\text{mid}}$ denote the values of the fields in
the middle region prior to gluing. Since the middle region is Calabi-Yau, we
have that $g_{\text{mid}}^{2}$ is a constant and $H_{\text{mid}}$ is zero. So
in other words, the contribution from the $M_{L}$ region is localized in
$M_{\text{mid}}$ near the locus $z_{\text{mid}}=0$. As we move to larger
values of $z_{\text{mid}}$, the profile of the dilaton will approach a
constant value. In the compact geometry, this normal direction corresponds to
moving further away from the marked point of the elliptic fiber where we
performed the gluing.

Turning next to the region $M_{L}$, we can again study the profile of the
dilaton and three-form flux. From our previous analysis, we know that $M_{L}$
is not Calabi-Yau, so there must be fluxes switched on. To track their
behavior near $D$, we introduce a normal coordinate $z_{L}$, and decompose the
Hermitian $(1,1)$-form $J$ as,
\begin{equation}
J=a_{\bot}J_{\bot}+a_{\Vert}J_{\Vert},
\end{equation}
where $J_{\Vert}$ are the components of the $(1,1)$ form along $D$, and
$J_{\bot}\sim\frac{i}{2}dz_{L}\wedge d\overline{z}_{L}$ is the contribution
normal to $D$. The $a$'s are position dependent contributions. We also know
from the F-theory description that the profile of the dilaton is, to leading
order, dependent on only $z_{L}$, the normal coordinates. So to leading order
these coefficients only depend on $z_{L}$.

Via the equations of the Strominger system, we see that
the dilaton is related to these coefficients as
\begin{equation}
\exp(2\phi_{L})\sim a_{\Vert}^{2},
\end{equation}
where there will be subleading contributions in the local geometry. The
$a_{\Vert}$ has dependence on the normal coordinates $z_{L}$. Turning next
to the profile of the three-form flux, we can now see that the three-form
flux equation of motion, $H=i(\overline{\partial}-\partial)J$ reduces to,
\begin{equation}
H\sim i(\overline{\partial}-\partial)a_{\Vert}\wedge J_{\Vert}.
\end{equation}
This has the general form of a flux which is concentrated near the divisor $D
$, and which spreads out in the direction normal to $D$. As we move to larger
values of $z_{L}$, this approximation breaks down, and we can see that most of
the $M_{L}$ region becomes a flux background with order one string coupling.

\subsection{Heterotic Localization \label{ssec:LOCALIZE}}

As we have already seen in the F-theory geometry, the seven-branes of
$X_{L,\text{mid}}$ are localized, and can remain dynamical even when gravity
is decoupled. For the proposed duality to hold, a similar localization must
happen in the heterotic configuration. Now, in contrast to F-theory, in
heterotic theory the perturbative gauge degrees of freedom come from a
ten-dimensional gauge field, so no localization would at first appear to be
possible.\footnote{Localization of gauge theory degrees of freedom can also
occur for heterotic strings compactified on a geometry with an orbifold
singularity. We emphasize that the mechanism we discuss in this section is not
of this type.} However, this implicitly assumes that the background value of
the dilaton is constant, an assumption we are violating in our proposed
duality. This behavior of the dilaton points to a localization mechanism for
the heterotic string.

To illustrate the main point, let us consider a simplified situation where we
take an abelian $D$-dimensional gauge theory, but with a position dependent
gauge coupling,%
\begin{equation}
S_{\text{kin}}=-\int\frac{1}{4g^{2}(x)}F\wedge\ast F.
\end{equation}
In physical terms, we see that if we hold fixed the background gauge coupling
$g(x)$, the fluctuations of the gauge field can become trapped on a subspace via the dielectric
effect proposed in \cite{Hooft, KogutSusskind, Kawai:2001kg, Ohta:2010fu}.
Indeed, the equations of motion reduce to
\begin{equation}
d\left(  \frac{1}{g^{2}(x)}\ast F\right)  = 0.
\end{equation}
One can further decompose the legs of the gauge field into directions along which the gradient of the
gauge coupling vanishes, and transverse directions along which the gradient of the gauge coupling does not vanish. The
transverse fluctuations of the gauge field are subject to a second order differential
equation, which in appropriate circumstances has an isolated massless mode \cite{Kawai:2001kg}.
One crude way to see this effect is to introduce a small infrared mass term for the gauge field.
Canonically normalizing the kinetic energy, we see that the position dependent mass
becomes big when the gauge coupling $g(x)$ is big. So in other words, the
gauge field is trapped in the regions of smaller gauge coupling.

We now see that a position dependent string coupling in heterotic theory should also lead
to localization of the gauge fields, as predicted by the F-theory geometry. The regions
of smaller coupling are those places where a gauge field becomes trapped. From this
perspective, there could in principle be many ways that the ten-dimensional gauge fields could become
localized on various subspaces.

Returning to the specific example encountered in our gluing construction,
recall that in the region $M_{\text{mid}}$, the heterotic string coupling is
position dependent, and given by,%
\begin{equation}
\frac{1}{g_{\text{het}}^{2}}=\frac{1}{g_{\text{mid}}^{2}}+f_{L}\text{.}%
\end{equation}
In the regime where we take $g_{\text{mid}}$ very large, we see that the
heterotic gauge field has become localized near the gluing region, with
falloff in the region $M_{\text{mid}}$ set by the profile of $f_{L}$. In the
more general situation where $g_{\text{mid}}$ is not arbitrarily large, we can
see that the heterotic gauge fields will still be localized, but that the
characteristic size will be set by a combination of $g_{\text{mid}}$ and
$f_{L}$.

Let us now turn to some preliminary aspects of how to go about finding vector
bundle solutions in these flux backgrounds. Our aim here is to simply sketch
the main aspects, we leave explicit examples to future work. To set up the
correspondence, we recall that when the elliptic fibration of the heterotic
geometry has a section, one can utilize the Fourier-Mukai transform to convert
a vector bundle on the $T^{2}$ fiber to a vector bundle on $M_{L}$. Of course,
the whole point of our construction is that in the compact geometry, before
deleting the divisor to reach $M_{L}$, the elliptic fibration need not have a
holomorphic section. Nevertheless, once we have deleted the subspace $P$ from the
base $S$ to reach $S_{L}$, as is necessary for the gluing construction anyway,
we do have a section, but at the expense of dealing with a non-compact
geometry. At a formal level, we can then relate this to the known results on
the construction of intersecting seven-branes with flux and their heterotic
duals. This follows the same procedure spelled out in eight dimensions in
\cite{Friedman:1997ih, Friedman:1997yq}, and its extension to four-dimensional
vacua (see for example \cite{BershadskyFOURD, Hayashi:2009ge, Andreas:2009uf,
DWIII, Marsano:2009gv}). The main subtlety is that in contrast to the standard
duality, $S_{L}$ is non-compact, so the topology of the bundle is not really
fixed until we glue back in $M_{L}~$to the geometry $M_{\text{mid}}$. Indeed,
we know that to get a dynamical gauge group, we need to glue into the ambient
geometries $X_{\text{mid}}$ and $M_{\text{mid}}$, respectively.

\section{Recoupling to Gravity \label{sec:GRAVITY}}

Our discussion so far has focussed on the building blocks necessary to realize
a non-compact version of heterotic/F-theory duality. Ultimately, we need to
recouple to gravity. This is accomplished by compactifying the middle region,
both in the F-theory geometry, as well as in the heterotic dual. In the
F-theory geometry this requires the appearance of at least three building
blocks, $X_{L}$, $X_{\text{mid}}$ and $X_{R}$ which we glue together to form a
compact F-theory geometry,%
\begin{equation}
X_{\text{F-th}}=X_{L}\cup_{Y_{L}}X_{\text{mid}}\cup_{Y_{R}}X_{R}.
\end{equation}
In a compact model, there will be some induced D3-brane charge. In the dual
F-theory description, this is satisfied by the tadpole constraint
\cite{Sethi:1996es}%
\begin{equation}
\frac{\chi(X_{\text{F-th}})}{24}=N_{D3}+\frac{1}{2}\underset{X_{\text{F-th}}%
}{\int}G_{4}\wedge G_{4},\label{tadpoles}%
\end{equation}
where $G_{4}$ is the four-form flux in the dual M-theory description and
$N_{D3}$ is the number of D3-branes.

\begin{figure}[ptb]
\begin{center}
\includegraphics[
height=1.4341in, width=5.2356in ]{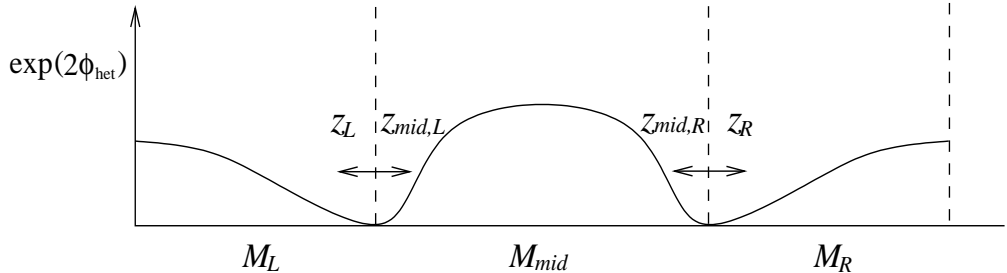}
\end{center}
\caption{Illustration of the heterotic dilaton profile in the different
regions. In the middle region $M_{\text{mid}}$ of the heterotic geometry, the
string coupling becomes large, localizing the gauge field degrees of freedom
in the gluing regions between $M_{L}$ and $M_{\text{mid}}$, and between
$M_{R}$ and $M_{\text{mid}}$.}%
\label{fig:configurations_2}%
\end{figure}

In the dual heterotic description, the compact geometry is given by gluing
together our non-compact dual building blocks
\begin{equation}
M_{\text{het}}=M_{L}\cup_{D_{L}}M_{\text{mid}}\cup_{D_{R}}M_{R}.
\end{equation}
Additionally, we need to piece together the profiles of the dilaton and
three-form flux. These modes have a non-trivial position dependence in $M_{L}$
and $M_{R}$, and asymptote to constant values deep in the region
$M_{\text{mid}}$. See figure \ref{fig:configurations_2} for a depiction of the
dilaton position dependence. In the heterotic theory, the analogue of the
tadpole constraint of line (\ref{tadpoles}) involves activating background
gauge field fluxes as well as NS5-branes wrapped on effective divisors.

Note that there is still just a single ten-dimensional vector bundle
$V=E_{8}\times E_{8}$ but that fluctuations become trapped in different
regions. As a consequence, the effective number of independent ten-dimensional
vector bundles increases. For example, if we assume that $M_{\text{mid}}$ is a
region of strong coupling, whereas $M_{L}$ and $M_{R}$ are perturbatively
realized, we see that the number of independent vector bundles will
effectively double. This is in accord with the behavior in the dual F-theory
geometry, where there are roughly speaking two independent $K3$ fibrations
which get glued together via $X_{\text{mid}}$.

Localization of the gauge theory degrees of freedom points to a number of
potential applications for model building. In the context of local F-theory
model building, breaking the GUT group involves activating a hypercharge flux.
This is possible in F-theory because the gauge fields of a seven-brane are
localized, and so can remain decoupled from the bulk axions which would
otherwise give a mass to the gauge fields \cite{BHVII, DWII}. This mechanism
has no analogue in heterotic Calabi-Yau compactification \cite{WittenU(1),
BHVII, DWII}.

But since we have now seen how to localize heterotic gauge fields, we should
expect a similar GUT\ breaking mechanism to hold in the heterotic string.
Indeed, for the generalized duality to hold true, this must be possible. Thus,
in addition to identifying a new physical mechanism for GUT\ breaking, this
will provide a useful check on our proposal.

Our plan in the following part of this section will be to elucidate how
heterotic hyperflux works in the presence of localized gauge fields. To this
end, we shall first review some features of bulk axion couplings to gauge
theory degrees of freedom. Then, we review the hyperflux mechanism for
F-theory compactification, and then translate this to our heterotic construction.

\subsection{Heterotic Hyperflux \label{ssec:HetHyper}}

In this subsection we exhibit a heterotic dual to the hyperflux mechanism. The
main idea is to show that an abelian flux $U(1)_{Y}\subset SU(5)_{GUT}$ can be
activated, but which also decouples from all bulk axions.

To frame our discussion, let us briefly review some aspects of the hyperflux
mechanism in F-theory \cite{BHVII, DWII} (see also \cite{BuicanVerlinde}). We
begin with F-theory compactified on a threefold base $B$, and study the worldvolume theory of a seven-brane
with gauge group $G$ wrapping $\mathbb{R}^{3,1}\times S$ for some K\"{a}hler
surface $S$. In the eight-dimensional gauge theory, we have the terms,%
\begin{equation}
S_{\text{10D}}\supset-M_{\ast}^{4}\underset{\mathbb{R}^{3,1}\times S}{\int
}\text{Tr}(F_{8D}\wedge\ast_{8}F_{8D})+\underset{\mathbb{R}^{3,1}\times
S}{\int}i^{\ast}(C_{4})\wedge\text{Tr}(F_{8D}\wedge F_{8D})+M_{\ast}%
^{6}\underset{\mathbb{R}^{3,1}\times B}{\int}dC_{4}\wedge\ast_{10}dC_{4},
\end{equation}
where $i^{\ast}(C_{4})$ is the pullback of the bulk four-form potential
$C_{4}$ onto $\mathbb{R}^{3,1}\times S$, $F_{8D}$ is the 8D\ field strength,
and $M_{\ast}$ is a characteristic UV scale where the large volume
approximation breaks down.

Suppose now we expand this theory around a non-trivial internal gauge field
flux valued in some abelian subgroup $U(1)\subset G$. For ease of exposition,
we treat all gauge fields as abelian. We decompose the form content of the
eight-dimensional field strength as,%
\begin{equation}
F_{8D}=F_{4D}+F_{S},
\end{equation}
for some non-zero background value of $F_{S}$. We also decompose the four-form
$C_{4}$ into a basis of internal harmonic two-forms on $B$,%
\begin{equation}
C_{4}=r^{\alpha}\wedge b_{\alpha},
\end{equation}
where $b_{\alpha}$ is a two-form on $B$, and $r^{\alpha}$ is a two-form on
$\mathbb{R}^{3,1}$ dual to an axion. Expanding around this background, we get
the four-dimensional terms,%
\begin{equation}
S_{\text{4D}}\supset-\frac{1}{4g_{U(1)}^{2}}\underset{\mathbb{R}^{3,1}}{\int
}F_{4D}\wedge\ast_{4}F_{4D}+\underset{\mathbb{R}^{3,1}}{\int}r^{\alpha}\wedge
F_{4D}\underset{S}{\int}i^{\ast}(b_{\alpha})\wedge F_{S}+M_{\ast}^{2}%
\underset{\mathbb{R}^{3,1}}{\int}dr^{\alpha}\wedge\ast_{4}dr_{\alpha}.
\end{equation}
The middle term is a coupling between an axion and a gauge field. When it is
non-zero, the abelian gauge field picks up a large mass of order $M_{\ast}$.

In F-theory GUTs, such couplings can be eliminated provided,%
\begin{equation}
\underset{S}{\int}i^{\ast}(b_{\alpha})\wedge F_{S}=0,
\end{equation}
for all harmonic two-forms $b_{\alpha}$ on $B$. This can be arranged by a
trivialization condition of the divisor dual to $F_{S}$ inside of $B$. The
embedding $i:S\rightarrow B$ induces the pullback map for cohomology,%
\begin{equation}
i^{\ast}:H^{2}(B)\rightarrow H^{2}(S).
\end{equation}
So, a nontrivial relative cohomology allows us to generate a hyperflux which
decouples from all bulk axions.

Now, in heterotic strings, this GUT breaking mechanism would at first appear
to be absent. As explained in \cite{WittenU(1)}, for heterotic strings
compactified on a Calabi-Yau threefold, the hyperflux mechanism is
unavailable. This is because of the interaction terms in the ten-dimensional
action,%
\begin{equation}
S_{\text{10D}}\supset-M_{\ast}^{6}\underset{\mathbb{R}^{3,1}\times M}{\int
}\frac{1}{g^{2}}\text{Tr}(F_{10D}\wedge\ast_{10}F_{10D})+\underset
{\mathbb{R}^{3,1}\times M}{\int}\left\vert d\Lambda+A\wedge F\right\vert ^{2},
\end{equation}
where $\Lambda$ is the two-form potential of the heterotic theory. Let us now
expand around a background value of the internal field strength $F_{M}$.
Decompose $\Lambda$ into a basis of harmonic two-forms $\lambda_{\alpha}$ on
$M$,%
\begin{equation}
\Lambda=c^{\alpha}\wedge\lambda_{\alpha},
\end{equation}
with $c^{\alpha}$ an axion of the four-dimensional theory. Then, upon
expanding with respect to an internal flux,%
\begin{equation}
F_{10D}=F_{4D}+F_{M},
\end{equation}
the four-dimensional effective action contains the terms,%
\begin{equation}
S_{\text{4D}}\supset-\frac{1}{4g_{U(1)}^{2}}\underset{\mathbb{R}^{3,1}}{\int
}F_{4D}\wedge\ast_{4}F_{4D}+\underset{\mathbb{R}^{3,1}}{\int}r^{\alpha}\wedge
F_{4D}\underset{M}{\int}\ast_{6}\lambda_{\alpha}\wedge F_{M}+M_{\ast}%
^{2}\underset{\mathbb{R}^{3,1}}{\int}dc^{\alpha}\wedge\ast_{4}dc_{\alpha
}\text{,}%
\end{equation}
where $r^{\alpha}$ is the two-form dual to the axion $c^{\alpha}$ in
four-dimensions. Again, the middle term is responsible for the St\"{u}ckelberg
mechanism of the four-dimensional effective theory. In the standard heterotic
compactification on a Calabi-Yau threefold, the harmonic two-forms
$\lambda_{\alpha}$ and $F_{M}$ are both representatives of elements in
$H^{2}(M)$, so the hyperflux mechanism is unavailable.

With a position dependent dilaton, however, we can localize the profile of the
heterotic gauge fields. It is therefore worth revisiting whether the hyperflux
mechanism holds in heterotic models. In fact, localization is by itself not
enough to ensure that a given heterotic gauge bundle configuration will
decouple from the axions. The main idea will be to formally construct a
non-trivial vector bundle on the \textquotedblleft standard\textquotedblright%
\ middle region $M_{\text{mid}}$, and then show that in the full geometry
$M_{\text{het}}$, it trivializes. In other words, we consider the embedding%
\begin{equation}
i:M_{\text{mid}}\rightarrow M_{\text{het}},
\end{equation}
and seek a non-trivial kernel to the pushforward%
\begin{equation}
i_{\ast}:H_{4}(M_{\text{mid}},%
\mathbb{Z}
)\rightarrow H_{4}(M_{\text{het}},%
\mathbb{Z}
).
\end{equation}
The localization of the ten-dimensional gauge fields near the gluing regions
$D_{L}$ and $D_{R}$ means that effectively, the GUT\ breaking flux is
localized on this lower-dimensional component of the geometry.

To construct examples of gauge field configurations which trivialize in the
full geometry, we can first construct a line bundle over $M_{\text{mid}}$
which, upon gluing, trivializes in the full geometry $M_{\text{het}}$. Along
these lines, recall that $M_{\text{mid}}$ is given by an elliptic fibration
with section over a base $D$. We shall assume that there are at least two
effective divisors $\sigma_{1},\sigma_{2}$ with homology classes $[\sigma
_{i}]\in H_{2}(D,\mathbb{Z)}$ such that $\sigma_{1}-\sigma_{2}$ is trivial
inside of $M_{L}$, but is non-trivial inside of $M_{\text{mid}}$. This can
happen because in $M_{\text{mid}}$ there is a section to the fibration, so
$\sigma_{1}$ and $\sigma_{2}$ lift to two non-trivial divisors $S_{1},S_{2}$
with homology classes $[S_{i}]\in H_{4}(M_{\text{mid}},\mathbb{Z)}$. So, let
us consider the line bundle $\mathcal{L}_{\text{mid}}=\mathcal{O}%
_{M_{\text{mid}}}(S_{1}-S_{2})$. Under the embedding map, we can pushforward
$\mathcal{L}_{\text{mid}} $ to a rank one sheaf on $M_{\text{het}}$. Observe,
however, that since $[S_{1}]=[S_{2}]$ in $H_{4}(M_{\text{het}},\mathbb{Z)}$,
that the topology of the line bundle is globally trivial, even though there is
a non-trivial flux localized along $D_{L}$ and $D_{R}$. Indeed, upon
restriction of $\mathcal{L}_{\text{mid}}$ to $D$, we get the line bundle
$\mathcal{O}_{D}(\sigma_{1}-\sigma_{2})$.

As consequence of this topological mechanism, all couplings to bulk axions
automatically vanish. This includes model-dependent axions coming from
harmonic two-forms of $M_{\text{het}}$, as well as the contribution from the
universal axion of a heterotic compactification. In our analysis, we have used
the gluing to the middle region as a means to track this possibility.
Following up on the discussion in subsection \ref{ssec:LOCALIZE}, it would be
quite interesting to understand this purely from the perspective of vector
bundles on $M_{L}$.

Finally, note that any holomorphic vector bundle on $M_{\text{mid}}$ which
trivializes in the full geometry will automatically define a consistent
solution to the Hermitian Yang-Mills equations. The reason is that the
Hermitian $(1,1)$ form $J_{mn}$ is a bulk mode defined over the entire
geometry $M_{\text{het}}$. So, there is automatically a representative flux
which satisfies the condition $F_{mn}J^{mn}=0$.

\section{Examples \label{sec:EXAMP}}

In this section we give some examples of how we expect our proposal to work.
First, we treat the specific case of F-theory with base $B=\mathbb{P}^{3}$.
Then, we present some generalizations.

\subsection{F-theory with $\mathbb{P}^{3}$ Base}

\begin{figure}
[ptb]
\begin{center}
\includegraphics[
height=1.3993in,
width=4.2635in
]%
{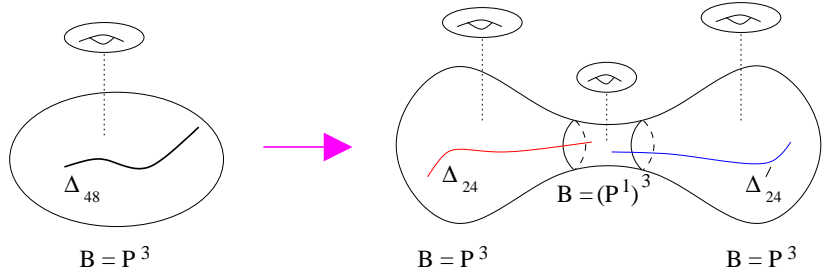}%
\caption{Depiction of a generalized stable degeneration limit for F-theory
with a $\mathbb{P}^{3}$ base. The discriminant locus of the model is specified
by a degree $48$ hypersurface in $\mathbb{P}^{3}$. In the stable degeneration
limit, we split this into two components $X_{L}$ and $X_{R}$ which are glued
together across an asymptotic cyclindrical region $X_{\text{mid}}$ which is
also Calabi-Yau. Each $X_{L}$ and $X_{R}$ carries a component of the
discriminant locus of degree $24$.}%
\label{p3example}%
\end{center}
\end{figure}

Let us consider the special case of F-theory on the base $B=\mathbb{P}^{3}$.
We recall that $\mathbb{P}^{3}$ is also the twistor space for $S^{4}$ via the
fibration,%
\begin{equation}
\mathbb{P}_{f}^{1}\rightarrow\mathbb{P}^{3}\rightarrow S^{4}\text{.}%
\end{equation}
At a given point of $S^{4}$, we can parameterize the sphere of complex
structures for the tangent space as the coset space $\mathbb{P}_{f}^{1}\simeq
SO(4)/U(2)$. An important feature of this fibration is that it does not admit
a global section and therefore is an excellent test case for our general
considerations. The homology ring for $\mathbb{P}^{3}$ is generated by the
hyperplane class $H$, and the canonical class for $B$ is $K_{B}=-4H$. The
twistor fiber $\mathbb{P}_{f}^{1}$ is a degree one curve in $\mathbb{P}^{3}$,
with class $H^{2}$. The absence of a section for the fibration means that the
$S^{4}$ does not exist as a four-manifold inside of $\mathbb{P}^{3}$. Our plan
will be to establish a stable degeneration for this geometry, as depicted in
figure \ref{p3example}, and then to use these building blocks to establish a
corresponding heterotic vacuum.%

\subsubsection{F-theory Building Blocks}

Now, although the $\mathbb{P}^{1}$-fibration for $\mathbb{P}^{3}$ has no
section, the elliptically fibered Calabi-Yau fourfold with $\mathbb{P}^{3}$
base does have a section. The minimal Weierstrass model in this case is,%
\begin{equation}
X_{\text{F-th}}=\left\{  y^{2}=x^{3}+f_{16}x+g_{24}\right\}  ,
\end{equation}
where $f_{16}$ and $g_{24}$ are respectively sections of $\mathcal{O}%
(-4K_{B})$ and $\mathcal{O}\left(  -6K_{B}\right)  $. Since $K_{\mathbb{P}%
^{3}}=-4H$, we have that $f$ and $g$ are degree $16$ and $24$ homogeneous
polynomials in variables $u_{1},...,u_{4}$ for $\mathbb{P}^{3}$. This example is rather
special, since it can be unfolded to just $I_{1}$ fibers \cite{Klemm:1996ts}.
The Hodge diamond for this fourfold is computed in Appendix A, and is (see
also \cite{Klemm:1996ts})
\begin{equation}%
\begin{array}
[c]{ccccc}%
1 & 0 & 0 & 0 & 1\\
0 & 3,878 & 0 & 2 & 0\\
0 & 0 & 15,564 & 0 & 0\\
0 & 2 & 0 & 3,878 & 0\\
1 & 0 & 0 & 0 & 1
\end{array}
,
\end{equation}
where the lower lefthand corner is $h^{0,0}(X_{\text{F-th}})$, and the upper
righthand corner is $h^{4,4}(X_{\text{F-th}})$. The topological invariants of
the Calabi-Yau and base include:%
\begin{equation}
\frac{\chi(X_{\text{F-th}})}{24}=972\text{, \ \ }c_{2}(X_{\text{F-th}}%
)c_{2}(X_{\text{F-th}})=8,256\text{, \ \ }\chi(B)=4\text{, \ \ }c_{1}%
(B)^{3}=64\text{, \ \ }c_{1}(B)c_{2}(B)=24.
\end{equation}

The high degree of the coefficients $f$ and $g$ reflects the fact that there
is no globally defined $K3$ fibration for this F-theory model. So it cannot
have a standard heterotic dual Calabi-Yau compactification. Indeed, since the
discriminant locus is a degree $48$ homogeneous polynomial, there are roughly
speaking two $K3$'s worth of gauge theory degrees of freedom.

However, since the base admits a $\mathbb{P}^{1}$ fibration (though one
without a holomorphic section), we expect some form of the duality to persist.
To construct the appropriate heterotic dual manifold, we now turn to the
F-theory building blocks. This will allow us to construct the corresponding
dual heterotic geometry.

As our first step, we need to take a stable degeneration limit of $X$ so that
it splits into two pieces $X_{L}$ and $X_{R}$ which are glued together along a
Calabi-Yau threefold $Y$. In general, this variety is singular along $Y$. Our
smoothing consists of three building blocks, each of which admits a $K3$
fibration,%
\begin{equation}
X=X_{L}\cup_{Y_{L}}X_{\text{mid}}\cup_{Y_{R}}X_{R}.
\end{equation}

So, let us begin by producing these building blocks by constructing an
elliptic fibration over $B$ which does admit a $K3$ fibration. To do this, we
modify the degrees of $f$ and $g$ to get a geometry $X_{L}^{\text{cpct}}$,
where the left building block is $X_{L}=X_{L}^{\text{cpct}}\backslash Y$. The
geometry $X_{L}^{\text{cpct}}$ is defined by%
\begin{equation}
X_{L}^{\text{cpct}}=\left\{  y^{2}=x^{3}+f_{8}x+g_{12}\right\}
,\label{XLequn}%
\end{equation}
for $f_{8}$ and $g_{12}$ sections of $\mathcal{O}_{B}(-2K_{B})$ and
$\mathcal{O}_{B}(-3K_{B})$. To verify that this defines a $K3$ fibration, we
consider the restriction of this model to the twistor $\mathbb{P}_{f}^{1}$.
Since the class of $\mathbb{P}_{f}^{1}$ is $H\cap H$, there is no change in
the degree upon restriction, and we get a $K3$ surface. In other words,
$X_{L}^{\text{cpct}}$ defines a $K3$ fibration over $S^{4}$. In Appendix A we
calculate some topological invariants of this geometry. The Hodge diamond for
$X_{L}^{\text{cpct}}$ is%
\begin{equation}%
\begin{array}
[c]{ccccc}%
0 & 0 & 0 & 0 & 1\\
0 & 370 & 0 & 2 & 0\\
0 & 0 & 1,702 & 0 & 0\\
0 & 2 & 0 & 370 & 0\\
1 & 0 & 0 & 0 & 0
\end{array}
,
\end{equation}
where the lower lefthand corner is $h^{0,0}(X_{L}^{\text{cpct}})$ as the upper
righthand corner is $h^{4,4}(X_{L}^{\text{cpct}})$. The Euler character of
$X_{L}^{\text{cpct}}$ is $\chi(X_{L}^{\text{cpct}})=2,448$.

Following our general procedure, we need to delete a Calabi-Yau threefold $Y$
from $X_{L}^{\text{cpct}}$ to produce a non-compact Calabi-Yau fourfold
$X_{L}=X_{L}^{\text{cpct}}\backslash Y$. In the base $B$, we need to pick an
appropriate effective divisor $D$ so that the restriction of the elliptic
fibration to $D$ defines an elliptic Calabi-Yau threefold. To begin, let us
recall that an elliptic fibration over $D$ will require $f$ and $g$ to
restrict to sections of $\mathcal{O}_{D}(-4K_{D})$ and $\mathcal{O}%
_{D}(-6K_{D})$, respectively. So in other words, we require%
\begin{equation}
\mathcal{O}_{B}(-2K_{B})|_{D}=\mathcal{O}_{D}(-4K_{D}). \label{restrictor}%
\end{equation}
On the other hand, we have, via the adjunction theorem,%
\begin{equation}
K_{D}=(K_{B}+[D])|_{D}=(n-4)H|_{D},
\end{equation}
where we have set $K_{B}=-4H$, and $D=nH$ for some $n>0$. So, we see that the
condition of equation (\ref{restrictor}) can be satisfied provided we take
$n=2$. We therefore conclude that the appropriate divisor is cut out by a
degree two homogeneous polynomial in $\mathbb{P}^{3}$, which generically
defines a $\mathbb{P}^{1}\times\mathbb{P}^{1}$.

The corresponding elliptic threefold $Y$ is then given in minimal Weierstrass
model by the presentation,%
\begin{equation}
Y=\left\{  y^{2}=x^{3}+f_{8,8}x+g_{12,12}\right\}  ,
\end{equation}
where $f_{8,8}$ denotes a polynomial of bidegree eight which is homogeneous in
the variables of each $\mathbb{P}^{1}$ factor of $D=\mathbb{P}^{1}%
\times\mathbb{P}^{1}$. The degree is fixed by the condition that $f_{8,8}$ be
a section of $-4K_{D}=8[\sigma_{1}]+8[\sigma_{2}]$, where the $[\sigma_{i}]$
are the divisor classes of the two $\mathbb{P}^{1}$ factors. Similarly,
$g_{12,12}$ denotes a section of $-6K_{D} = 12[\sigma_{1}]+ 12[\sigma_{2}]$.
Note that each $\mathbb{P}^{1}$ factor defines the base of a $K3$ fibration
for $Y$.

Having introduced the left region $X_{L}$, we can now construct the middle
region $X_{\text{mid}}$. Prior to gluing, this is given by a Calabi-Yau
fourfold with base $D\times\mathbb{P}^{1}$, that is, we have F-theory on the
threefold base $\mathbb{P}^{1}\times\mathbb{P}^{1}\times\mathbb{P}^{1}$. Each
pair of $\mathbb{P}^{1}$ factors of the base defines the base of a $K3$
fibration for $X_{\text{mid}}$. The elliptic model in this case is%
\begin{equation}
X_{\text{mid}}=\left\{  y^{2}=x^{3}+f_{8,8,8}x+g_{12,12,12}\right\}  ,
\end{equation}
in the obvious notation. The Euler character of $X_{\text{mid}}$ is
$\chi(X_{\text{mid}})=17,568$. For additional details on this fourfold, see
for example \cite{Klemm:1996ts}.

In a similar way, we can construct $X_{R}$ and glue it to $X_{\text{mid}}$.
The full compact geometry is then
\begin{equation}
X_{\text{F-th}}=X_{L}\cup_{Y_{L}}X_{\text{mid}}\cup_{Y_{R}}X_{R}.
\end{equation}
The discriminant loci of $X_{L}$ and $X_{R}$ are each of degree
$24$, that is, they each give a $K3$'s worth of seven-branes. The discriminant
locus of the middle region $X_{\text{mid}}$ is really the completion of the
left and right regions, and so does not contribute additional independent
seven-brane gauge theory degrees of freedom.

A special feature of this example is that we can
also consider a more general base threefold in the middle
region given by fibering $D$ over $\mathbb{P}^{1}$, that is, by specifying a
non-trivial $\mathbb{P}_{(1)}^{1}\times\mathbb{P}_{(2)}^{1}$ bundle over the
base $\mathbb{P}^{1}$. There are then two integers we can specify,
corresponding to the choice of Hirzebruch surfaces $\mathbb{P}_{(1)}%
^{1}\rightarrow\mathbb{P}^{1}$ and $\mathbb{P}_{(2)}^{1}\rightarrow
\mathbb{P}^{1}$, of respective degrees $k_{1}$ and $k_{2}$. To be consistent
with the existence of a crepant resolution of the elliptic model, we need to
take $-12\leq k_{i}\leq12$. In the dual heterotic configuration, this choice
of $k_{i}$ will correspond to a choice of instanton number in an $E_{8}$
factor: In the left region $M_{L}$ we have instanton numbers $(12-k_{1}%
,12-k_{2})$, while in the right region $M_{R}$ we have instanton numbers
$(12+k_{1},12+k_{2})$.

\subsubsection{Heterotic Building Blocks}

Now that we have stated the F-theory building blocks, we turn to the heterotic
dual geometry. Following our proposal, we replace the $\mathbb{P}_{f}^{1}$
twistor fiber by an elliptic curve. To begin, let us consider $M_{L}$, the
dual for the building block $X_{L}$. The non-compact component $M_{L}$ is
reached by deleting a divisor from the compact $M_{L}^{\text{cpct}} $, which
is the $T^{2}~$fibration over $S$.~To replace the $\mathbb{P}_{f}^{1}$ fiber
by an elliptic fiber $\mathbb{E}_{f}$, we mark four points on $\mathbb{P}%
_{f}^{1}$ to define a double cover $\mathbb{E}_{f}\twoheadrightarrow
\mathbb{P}_{f}^{1}$. We extend this to a double cover of the threefold base
$B$ as follows. The four points for the branched cover are fixed by
intersecting the twistor fiber with a $K3$ divisor $C$ with divisor class
$[C]=4H$ in the $\mathbb{P}^{3}$. We can therefore define a double cover,%
\begin{equation}
\omega:\widehat{B}\rightarrow B
\end{equation}
which is branched over this $K3$ surface. We propose to take $M_{L}^{\text{cpct}} = \widehat{B}$.

Let us now describe the double cover $\widehat{B}$ in more detail. We
construct this by introducing a line bundle $\mathcal{L}=\mathcal{O}(2H)$ on
$\mathbb{P}^{3}$ so that our $K3$ surface $C$ is a zero section of
$\mathcal{L}^{\otimes2}$. Then, the canonical class of $\widehat{B}$ is given
by,%
\begin{equation}
K_{\widehat{B}}=\omega^{\ast}(K_{B}\otimes\mathcal{L}),
\end{equation}
so that the canonical class $K_{\widehat{B}}=-2\widehat{H}$, where the
homology ring is generated by $\widehat{H}$, the pullback of the hyperplane
class, which satisfies the relation $\widehat{H}^{3}=2$. For example, in the
upstairs geometry, the divisor class for a $K3$ surface is $2\widehat{H}$, and
the divisor class for a $\mathbb{P}^{1}\times\mathbb{P}^{1}$ is $\widehat{H}$.
The pushforward map,%
\begin{equation}
\omega_{\ast}:H_{\bullet}(\widehat{B},\mathbb{Z})\rightarrow H_{\bullet
}(B,\mathbb{Z})
\end{equation}
sends $\widehat{H}\mapsto H$.

Let us next describe some topological properties of $\widehat{B}$. The Euler
character of the new threefold $\widehat{B}$ is fixed by the standard formula
for a double cover branched along $C$ to be,%
\begin{equation}
\chi(\widehat{B})=2\chi(B)-\chi(C)=-16\text{,}%
\end{equation}
that is, we have two-fold cover of $B$, and we have deleted $C$, and glued it
back in once. Additionally, the new threefold has non-vanishing Hodge numbers,%
\begin{align}
h^{i,i}(\widehat{B}) &  =1\text{ for }i=0,...,3,\\
h^{2,1}(\widehat{B}) &  =h^{1,2}(\widehat{B})=10.
\end{align}
Finally, we can compute the Chern classes of $\widehat{B}$. The double cover
for this example can be described by a degree four hypersurface in the
weighted projective space $\mathbb{P}_{[1^{4},2]}^{4}$. Letting $H_{W}$ denote
the hyperplane class of the ambient weighted projective space, the divisor
class of $\widehat{B}$ is $[\widehat{B}]=4H_{W}$. The Chern class for
$\widehat{B}$ follows from expanding to third order in the divisor class,
\begin{equation}
c(\widehat{B})=\frac{(1+H_{W})^{4}(1+2H_{W})}{(1+4H_{W})}=1+2H_{W}+6H_{W}%
^{2}-8H_{W}^{3}.
\end{equation}
In the ambient space, we have $H_{W}^{4}=1/2$, so we can extract the numerical
invariants,%
\begin{equation}
c_{1}(\widehat{B})^{3}=16\text{, \ \ }c_{1}(\widehat{B})c_{2}(\widehat
{B})=24\text{, \ \ }c_{3}(\widehat{B})=-16.
\end{equation}

Now, as it stands, $M_{L}$ only defines a part of the full heterotic geometry.
Taking our cue from the F-theory geometry, we need to glue this into a
geometry $M_{\text{mid}}$ dual to $X_{\text{mid}}$. On the F-theory side of
the correspondence, the middle Calabi-Yau fourfold is given by a $K3$
fibration over $D=\mathbb{P}^{1}\times\mathbb{P}^{1}$. So, the appropriate
heterotic dual geometry in the middle is the elliptically fibered Calabi-Yau
threefold,%
\begin{equation}
M_{\text{mid}} = Y = \left\{  y^{2}=x^{3}+f_{8,8}x+g_{12,12}\right\}  .
\end{equation}
The Hodge numbers and Euler character for this threefold are
\begin{equation}
h^{1,1}(M_{\text{mid}})=3,\text{ \ \ }h^{2,1}(M_{\text{mid}})=243\text{,
\ \ }\chi(M_{\text{mid}})=-480.
\end{equation}
Finally, the heterotic dual for $X_{R}$ is $M_{R}$, that is, another copy of
$M_{L}$, and is constructed in the same way as $M_{L}$. In the heterotic
theory, these geometric building blocks are then glued together to construct
the full compact six-manifold,%
\begin{equation}
M_{\text{het}}=M_{L}\cup_{D_{L}}M_{\text{mid}}\cup_{D_{R}}M_{R}\text{.}%
\end{equation}

To complete the analysis, we also need to specify the profile of the heterotic
fields on the other side of the duality. Here, we must again take our guidance
from the F-theory geometry. First of all, deep in the middle region
$M_{\text{mid}}$, we have a standard compactification of heterotic strings on
a Calabi-Yau threefold. This means the heterotic dilaton can be taken to be a
constant, and there is no three-form flux switched on. A particularly
interesting feature of this specific heterotic dual is that the presence of
more than one $K3$ fibration in the F-theory geometry means we have various
string/string dualities in the heterotic theory.

Now, as we move closer to the gluing regions, the curvature of the metric
becomes more pronounced. Additionally, we can see that the profile of the
dilaton as well as the three-form flux also changes. Near the gluing locus
$D_{L}$, we see in particular that the profile of the string coupling becomes
weakly coupled, while it can be bigger deep in the $M_{L}$ and $M_{\text{mid}%
}$ regions. This enforces the localization of the heterotic gauge fields near
the gluing region, which is simply the heterotic dual of the familiar
localization of gauge theory degrees of freedom in the F-theory geometry.

Finally, deep in the regions $M_{L}$ and $M_{R}$, we can see that fluxes must
be switched on. The simplest way to see this is to observe that even after
deleting $D_{L}$ to reach $M_{L}$, we still have a non-compact positive
curvature six-manifold. Indeed, to reach a non-compact Calabi-Yau threefold,
we would have needed to delete a $K3$ surface. It is beyond the scope of the
present work to find an explicit solution to the metric and background fluxes
in this region, though we can see that the duality with F-theory clearly
predicts the existence of such a solution.

\subsubsection{Hyperflux}

One check of the duality we can already perform involves the construction of a
heterotic hyperflux. In the F-theory model, suppose we have a seven-brane
wrapping a divisor $\mathbb{P}^{1}\times\mathbb{P}^{1}$ in $\mathbb{P}^{3}$.
There is a single generator $H$ for the homology ring of $\mathbb{P}^{3}$
whereas there are two generators $\sigma_{1}$ and $\sigma_{2}$ for
$\mathbb{P}^{1}\times\mathbb{P}^{1}$. Indeed, the two-cycle
$\sigma_{1}-\sigma_{2}$ trivializes in $\mathbb{P}^{3}$. The seven-brane
two-form flux Poincar\'{e} dual to this class gives a configuration which
decouples from the bulk axions.

We can now see how a similar mechanism operates in the heterotic dual
configuration. Let us return to our discussion in subsection
\ref{ssec:HetHyper}. There, we showed how to build up a heterotic gauge field
configuration which breaks $SU(5)_{\text{GUT}}$ to the Standard Model gauge
group by activating a flux in the $U(1)_{Y}\subset SU(5)_{\text{GUT}}$
subgroup. First, we construct a line bundle on $M_{\text{mid}}$ given by
\begin{equation}
\mathcal{L}_{\text{mid}}=\mathcal{O}_{M_{\text{mid}}}(S_{1}-S_{2}),
\end{equation}
where $S_{i}$ are the divisor classes coming from the two $K3$ fibers of the
elliptic fibration $T^{2}\rightarrow M_{\text{mid}}\mathbb{\rightarrow
P}_{(1)}^{1}\times\mathbb{P}_{(2)}^{1}$. Upon restriction to the base
$D=\mathbb{P}_{(1)}^{1}\times\mathbb{P}_{(2)}^{1}$, the line bundle becomes
\begin{equation}
\mathcal{L}_{\text{mid}}|_{D}=\mathcal{O}_{D}(\sigma_{1}-\sigma_{2}),
\end{equation}
where $\sigma_{i}$ is the divisor class for one of the $\mathbb{P}_{(i)}^{1}$
factors. The important feature is that this class $\sigma_{1}-\sigma_{2}$
trivializes in $M_{\text{het}}$.

\subsection{Generalizations}

In this subsection we briefly discuss some possible generalizations to other
geometries. As we have repeatedly emphasized, the main feature of our proposed
duality is that the F-theory threefold base needs to have a $\mathbb{P}^{1}$
fibration over a four-manifold $S$. This fibration need not have a section,
and indeed, to realize the hyperflux mechanism, it seems necessary to require
the absence of a section.

One way to arrange examples of such six-manifolds is to take the twistor space
of a four-manifold. For example, the twistor space for $S^{4}$ is
$\mathbb{P}^{3}$, and the twistor space for $\mathbb{P}^{2}$ is the flag
manifold,%
\begin{equation}
\text{Tw}(\mathbb{P}^{2})=\mathbb{F}_{1,2,3}=\frac{U(3)}{U(1)\times U(1)\times
U(1)},
\end{equation}
which can also be presented as a bidegree one hypersurface in $\mathbb{P}%
_{(1)}^{2}\times\mathbb{P}_{(2)}^{2}$. This shares many of the properties of
the $\mathbb{P}^{3}$ base example, and after performing a stable degeneration
limit, provides another class of heterotic/F-theory building blocks.

The \textit{compact} four-manifolds which produce a K\"{a}hler twistor space
are conformal to either $S^{4}$ or $\mathbb{P}^{2}$ \cite{HitchinTwistor}. For
other compact four-manifolds, the twistor space is not K\"{a}hler.
For example, the twistor space for $\mathbb{P}^{1}\times\mathbb{P}^{1}$ is
complex, but not K\"{a}hler. Even so, we expect that our building block
construction would still apply. The reason is that to perform the gluing, we
need to delete a subspace from the twistor space. This deleting then allows us
to setup a globally defined non-compact K\"{a}hler threefold. So, the
individual building blocks for the F-theory geometry still involve non-compact
threefold bases glued along appropriate subspaces. On the heterotic side of
the construction, the duality will involve a branched double cover of the
non-compact twistor space. Here, some of the gluing data is also packaged in
terms of the profile of the heterotic fields. It would be quite interesting to
give explicit examples along these lines.

In six-dimensional F-theory vacua, another generalization is to consider the
Hirzebruch surfaces, that is, by taking a more general choice of
$\mathbb{P}^{1}$ fibration over a base $\mathbb{P}^{1}$. In a similar spirit,
one can consider the case of other $\mathbb{P}^1$ fibrations over a base $S$. Just as in the
Hirzebruch examples (see for example \cite{MorrisonVafaI, MorrisonVafaII,
BershadskyPLUS}), we expect there to be non-trivial restrictions on the degree
to be compatible with the existence of a smooth resolution for the elliptic fibration.

Much of our discussion has focussed on the simplest case of three building
blocks, for example $X_{L}$, $X_{\text{mid}}$ and $X_{R}$ in the F-theory
geometry. It would also be interesting to contemplate the case of
more building blocks perhaps in the spirit of
\cite{Donagi:2012ts}. There is a subtlety here, however,
because the construction of a complete metric allows us to cut out at most one
disjoint divisor from a positive curvature space such as $X_{L}^{\text{cpct}}$.

It would also be interesting to see whether there is a characterization
of the heterotic compactification as an information
geometry. Indeed, following up on the approach to string compactification
proposed in \cite{Heckman:2013kza}, the existence of our proposed duality means that
a low energy four-dimensional observer should not be able to distinguish the two spaces.

\newpage

\section{Conclusions \label{sec:CONCLUDE}}

In this note we have proposed a generalization of heterotic/F-theory duality.
On the F-theory side, the building blocks of the duality are non-compact
elliptically fibered Calabi-Yau fourfolds which also admit a
$K3$ fibration. These are glued together to form a compact elliptic Calabi-Yau
fourfold which need not have a global $K3$ fibration. On the heterotic side, the
$K3$ fiber of each F-theory building block is replaced by a $T^{2}$ fiber. In the
heterotic description, the gluing also involves a non-trivial three-form flux
and position dependent dilaton. Using our proposal, we reach new compact
examples of heterotic/F-theory duality pairs. This leads to a localization of
heterotic gauge field degrees of freedom in various regions of the geometry,
and also provides a heterotic version of the hyperflux mechanism for breaking
GUT\ groups. In other words, we have used F-theory to argue for the existence
of a new class of heterotic flux vacua. In the remainder of this section we
discuss some additional avenues of investigation.

In this work we have mainly focussed on the general contours of our proposal,
emphasizing in particular the simple form of the geometric F-theory building
blocks. It would clearly be useful to confirm in purely heterotic terms the
exact form of the background fields necessary to solve the equations of
motion. Along these lines, it would be important to verify that the resulting
low energy effective action defined by the heterotic compactification indeed
matches to the one defined by the F-theory model. In the case of heterotic
compactification on a model with a large radius limit, there is a simple
topological check which can be performed \cite{Grimm:2012yq}. It would be
interesting to extend this analysis to the class of flux vacua considered here.

On the other hand, one might instead take the F-theory geometry as a
definition of what a generalized heterotic vacuum ought to be. From this
perspective, the relevant issue is to demonstrate existence of a solution and
its topology rather than a direct construction of all background fields.

Along these lines, one ingredient which would be very interesting to work out
in more detail concerns the construction of heterotic vector bundles on
branched covers of twistor space. Roughly speaking, our proposal points to a
generalization of the standard spectral cover construction which should hold
even when the elliptic fibration of the heterotic model does not possess a
holomorphic section. Another generalization concerns giving a heterotic dual
description of T-branes (see for example \cite{TBRANES, glueI, glueII,
Anderson:2013rka}) for such flux vacua. Expanding on these details further would be
most interesting.

Finally, though we used the F-theory dual to motivate the existence of a
heterotic hyperflux mechanism, it should be possible to realize examples of
heterotic hyperflux even if there is no F-theory dual. Compared with
standard Calabi-Yau compactification, the main ingredient we have identified
is a position dependent dilaton profile to trap the 10D gauge fields on
regions of the geometry, and the existence of vector bundles which are
non-trivial on components of a gluing construction, but which are globally
trivial. This points to a potentially vast generalization of heterotic model building.

\section*{Acknowledgements}

JJH thanks K. Becker and S. Katz for many helpful discussions and
collaboration at an early stage of this work. We thank L.B. Anderson and C.
Vafa for helpful discussions and comments on an earlier draft. We also thank
M. Alim, B. Haghighat, M. Esole, E. Scheidegger, W. Taylor, and E. Witten for helpful discussions. The work
of JJH is supported by NSF grant PHY-1067976. The work of H. Lin and S.-T. Yau
is supported by NSF grant DMS-1159412, NSF grant PHY-0937443, NSF grant
DMS-0804454, and in part by the Fundamental Laws Initiative of the Center for
the Fundamental Laws of Nature, Harvard University.


\appendix

\section{Elliptic Fourfolds with a $\mathbb{P}^{3}$\ Base}

In this Appendix we collect some properties of elliptically fibered fourfolds
over a $\mathbb{P}^{3}$ base. We assume the fourfold embeds in a
$\mathbb{P}_{[1,2,3]}^{2}$ bundle over $\mathbb{P}^{3}$. Let $[z:y:x]$ denote
the coordinates of the weighted projective space $\mathbb{P}_{[1,2,3]}^{2}$
and $[u_{1}:u_{2}:u_{3}:u_{4}]$ for $\mathbb{P}^{3}$. The minimal Weierstrass
model is of the familiar form
\begin{equation}
X=\left\{  y^{2}=x^{3}+f_{4N}(u)xz^{4}+g_{6N}(u)z^{6}\right\}  ,
\label{hypersurface}%
\end{equation}
where $f_{4N}$ and $g_{6N}$ are homogeneous polynomials of degrees $4N$ and
$6N$, respectively. To extract the intersection theory of the fourfold, we
follow the same methodology reviewed for example in \cite{Klemm:1996ts,
DenefReview}.

We view our fourfold as a hypersurface in the toric variety defined by the
gauged linear sigma model with variables $u_{i}$, $x,y,z$ and $U(1)\times
U(1)$ charge assignments
\begin{equation}%
\begin{tabular}
[c]{|l|l|l|l|l|l|l|l|}\hline
& $u_{1}$ & $u_{2}$ & $u_{3}$ & $u_{4}$ & $x$ & $y$ & $z$\\\hline
$U(1)_{1}$ & $1$ & $1$ & $1$ & $1$ & $0$ & $0$ & $-N$\\\hline
$U(1)_{2}$ & $0$ & $0$ & $0$ & $0$ & $2$ & $3$ & $1$\\\hline
\end{tabular}
\ \ \ \ .
\end{equation}
We ignore the orbifold singularities of the fiber since the hypersurface
avoids these points anyway. Let $D_{i}$ and $D_{z}$ denote the divisor classes
for $u_{i}=0$, and $z=0$, respectively. Introduce the divisor classes,%
\begin{equation}
\Sigma\equiv\lbrack D_{1}]\text{ \ \ and \ \ }F=[D_{z}]+N[D_{1}].
\end{equation}
The intersection numbers for the divisors satisfy
\begin{equation}
\Sigma^{5}=0\text{, \ \ }\Sigma^{4}F=0\text{, \ \ }\Sigma^{3}F^{2}=1/6\text{,
\ \ }\Sigma^{2}F^{3}=N/6\text{, \ \ }\Sigma F^{4}=N^{2}/6\text{, \ \ }%
F^{5}=N^{3}/6.
\end{equation}
The fractional numbers are due to the orbifold singularity of the weighted
projective bundle.

Let us now turn to the intersection theory of the fourfold $X$, which is a
hypersurface in this ambient toric variety. The divisor class for the
hypersurface of line (\ref{hypersurface}) is $[X]=6F$, and the Chern classes
for $X$ are obtained by applying the splitting theorem
\begin{equation}
c(X)=\frac{(1+\Sigma)^{4}(1+2F)(1+3F)(1+F-N\Sigma)}{(1+6F)},
\end{equation}
which we expand to fourth order in the divisor classes.
Computing all intersection theoretic formulae in the ambient toric variety, we
extract the integrated Chern classes
\begin{align}
c_{1}^{4}  &  =0\text{, \ \ }c_{1}^{2}c_{2}=12N(N-4)^{2}\text{, \ \ }%
c_{1}c_{3}=24N(3N-2)(N-4),\\
c_{4}  &  =72N(6N^{2}-4N+1)\text{, \ \ }c_{2}^{2}=48N(3N^{2}-2N+3).
\end{align}

Using these invariants, we can extract the values of the indices%
\begin{equation}
\chi_{q}(X)=\underset{p}{\sum}(-1)^{p}h^{0,p}(X,\Omega_{X}^{q,0}),\text{
\ \ and \ \ }\chi(X)=\underset{i}{\sum}(-1)^{i}h^{i}(X),
\end{equation}
which in terms of the integrated Chern classes are
\begin{align}
\chi_{0}  &  =\frac{1}{720}\left(  -c_{4}+c_{1}c_{3}+3c_{2}^{2}+4c_{1}%
^{2}c_{2}-c_{1}^{4}\right) ,\\
\chi_{1}  &  =\frac{1}{180}\left(  -31c_{4}-14c_{1}c_{3}+3c_{2}^{2}+4c_{1}%
^{2}c_{2}-c_{1}^{4}\right) ,\\
\chi_{2}  &  =\frac{1}{120}\left(  79c_{4}-19c_{1}c_{3}+3c_{2}^{2}+4c_{1}%
^{2}c_{2}-c_{1}^{4}\right) .
\end{align}
The specific values for our example are,
\begin{equation}
\chi_{q}(X)=\underset{p}{\sum}(-1)^{p}h^{0,p}(X,\Omega_{X}^{q,0}),\text{
\ \ and \ \ }\chi(X)=\underset{i}{\sum}(-1)^{i}h^{i}(X),
\end{equation}
which in terms of the integrated Chern classes are
\begin{align}
\chi_{0}  &  =\frac{1}{6}(N-1)N(N+1)-N(N-2),\\
\chi_{1}  &  =-\frac{232}{3}(N-1)N(N+1)+2N(36N-49),\\
\chi_{2}  &  = (277N^{2}-142N+27)  N,\\
\chi &  =72N(6N^{2}-4N+1).
\end{align}
where we have also included the Euler character of the fourfold.

The cases of interest to us in this paper are $N=4$ and $N=2$. The former case
is a Calabi-Yau fourfold which we have referred to as $X_{\text{F-th}}$. The
latter case is the building block $X_{L}^{\text{cpct}}$, which has positive
curvature. From our index formulae, we can extract the specific values of the
numerical invariants as well as the Hodge numbers for both cases.

For $N=4$, we have
\begin{align}
c_{1}(X_{\text{F-th}})^{4}  &  =0\text{, \ \ }c_{1}(X_{\text{F-th}})^{2}%
c_{2}(X_{\text{F-th}})=0\text{, \ \ }c_{1}(X_{\text{F-th}})c_{3}%
(X_{\text{F-th}})=0,\\
c_{4}(X_{\text{F-th}})  &  =23,328\text{, \ \ }c_{2}(X_{\text{F-th}}%
)^{2}=8,256,\\
\chi_{0}(X_{\text{F-th}})  &  =2\text{, \ \ }\chi_{1}(X_{\text{F-th}%
})=-3,880\text{, \ \ }\chi_{2}(X_{\text{F-th}})=15,564,
\end{align}
and the Hodge numbers are
\begin{equation}%
\begin{array}
[c]{ccccc}%
1 & 0 & 0 & 0 & 1\\
0 & 3,878 & 0 & 2 & 0\\
0 & 0 & 15,564 & 0 & 0\\
0 & 2 & 0 & 3,878 & 0\\
1 & 0 & 0 & 0 & 1
\end{array}
,
\end{equation}
where the lower lefthand corner is $h^{0,0}(X_{\text{F-th}})$, and the upper
righthand corner is $h^{4,4}(X_{\text{F-th}})$.

For $N=2$, we have
\begin{align}
c_{1}(X_{L}^{\text{cpct}})^{4}  &  =0\text{, \ \ }c_{1}(X_{L}^{\text{cpct}%
})^{2}c_{2}(X_{L}^{\text{cpct}})=96\text{, \ \ }c_{1}(X_{L}^{\text{cpct}%
})c_{3}(X_{L}^{\text{cpct}})=-384,\\
c_{4}(X_{L}^{\text{cpct}})  &  =2,448\text{, \ \ }c_{2}(X_{L}^{\text{cpct}%
})^{2}=1,056,\\
\chi_{0}(X_{L}^{\text{cpct}})  &  =1\text{, \ \ }\chi_{1}(X_{L}^{\text{cpct}%
})=-372\text{, \ \ }\chi_{2}(X_{L}^{\text{cpct}})=1,702,
\end{align}
and the Hodge numbers are
\begin{equation}%
\begin{array}
[c]{ccccc}%
0 & 0 & 0 & 0 & 1\\
0 & 370 & 0 & 2 & 0\\
0 & 0 & 1,702 & 0 & 0\\
0 & 2 & 0 & 370 & 0\\
1 & 0 & 0 & 0 & 0
\end{array}
,
\end{equation}
where the lower lefthand corner is $h^{0,0}(X_{L}^{\text{cpct}})$, and the
upper righthand corner is $h^{4,4}(X_{L}^{\text{cpct}})$.

\newpage

\bibliographystyle{utphys}
\bibliography{HyperFlux}

\end{document}